\documentclass[12pt]{article}
\usepackage{epsf}
\title{ {\bf
Unparticle-two scalars mixing beyond the standard model and the
muon anomalous magnetic moment}}
\author{\vspace{1cm}\\
        {\bf E. O. Iltan}
        \thanks{E-mail address:
        eiltan@newton.physics.metu.edu.tr}
\\
Physics Department, Middle East Technical University \\
        Ankara, Turkey \\
        }

\date{}

\begin{document}
\setlength{\baselineskip}{24pt}
\maketitle
\setlength{\baselineskip}{7mm}
\begin{abstract}
We study the the electroweak symmetry breaking induced by the
interaction of unparticles  and the SM Higgs with an additional
complex scalar. Furthermore, we estimate the contribution of the
mixing scalars on the muon anomalous magnetic moment. We observe
that this contribution is at least two order less than the
discrepancy between the experimental and the SM results of the
muon anomalous magnetic moment.
\end{abstract}
\thispagestyle{empty}
\newpage
\setcounter{page}{1}
%%%
%%%
%\section{Introduction}
%
The electroweak symmetry breaking is an interesting phenomena
which has not understood yet. A possible hidden sector beyond the
standard model (SM) can be among the candidates to explain this
breaking. Such a hidden sector has been proposed by Georgi
\cite{Georgi1,Georgi2} as a hypothetical scale invariant one with
non-trivial infrared fixed point. It is beyond the SM at high
energy level and it comes out as new degrees of freedom, called
unparticles, being massless and having non integral scaling
dimension $d_u$. The interactions of unparticles with the SM
fields in the low energy level is defined by the effective
lagrangian (see for example \cite{SChen}).

Recently, the possibility of the electroweak symmetry breaking
from unparticles has been introduced \cite{JPLee} (see also
\cite{Kikuchi2} which leads to the origin of the electroweak
symmetry breaking in the unparticle physics, that is caused by the
mixing between the unparticle and the Higgs boson.). The idea is
based on the interaction of the SM scalar sector with the
unparticle operator in the form
$\lambda\,(\Phi^\dagger\,\Phi)\,O_U$ where $\Phi$ is the SM scalar
and $O_U$ is the unparticle operator with mass dimension $d_u$
(see \cite{Rajaraman, ADelgado, ADelgado2, Feng, Kikuchi,
RZwicky}). Since unparticles look like a number of $d_u$ massless
particles with mass dimension one, the operator $O_u$ can be
considered in the form of $(\phi^*\,\phi)^\frac{d_u}{2}$ which
induces the interaction term
\begin{eqnarray}
V\sim\lambda\,(\Phi^\dagger\,\Phi)\,(\phi^*\,\phi)^\frac{d_u}{2}\,,
\label{Vint}
\end{eqnarray}
driving the electroweak symmetry breaking in the tree level
\cite{JPLee}\footnote{See \cite{Weinberg} for the necessity of the
radiative corrections for the electroweak symmetry breaking from
hidden sector due to the interaction in the form
$\lambda\,(\Phi^\dagger\,\Phi)\,\phi^*\,\phi$.}.

In the present work we introduce a toy model that we extend the
scalar sector by considering a shadow Higgs one, the complex
scalar $\phi_2$ (see \cite{WFChang})\footnote{Here the $U(1)_s$
invariant Lagrangian including the shadow sector and the SM one
reads
\begin{eqnarray}
L=L_{SM}-\frac{1}{4}\,X^{\mu\nu}\,X_{\mu\nu}+|\Big(\partial_\mu-
\frac{1}{2}\,g_s\,X_\mu\Big)\,\phi_2|^2- V(\Phi_1, \phi_2,
\phi)\nonumber
\end{eqnarray}
where $g_s$ is the gauge coupling of $U(1)_s$.} in addition to the
the SM Higgs and we study the mechanism for the electroweak
symmetry breaking due to the unparticle-neutral scalars mixing, by
following the procedure given in \cite{JPLee}. Furthermore, we
estimate the contribution of the mixing scalar spectrum on the
anomalous magnetic moment (AMM) of the muon. Here, we expect that
this contribution is stronger compared to the case that only one
Higgs doublet exists and mixes with the unparticle sector.  Now,
we will construct the toy model by starting with the scalar
potential which is responsible for the unparticle-neutral scalars
mixing:
\begin{eqnarray}
V(\Phi_1, \phi_2, \phi)&=& \lambda_0 (\Phi_1^\dagger\,
\Phi_1)^2+\lambda^\prime_0 (\phi_2^*\, \phi_2)^2+\lambda_1
(\phi^*\, \phi)^2 \nonumber \\ &+& 2 \lambda_2\,\mu^{2-d_u}\,
(\Phi_1^\dagger\, \Phi_1)\,(\phi^*\, \phi)^\frac{d_u}{2}+2
\lambda^\prime_2\,\mu^{2-d_u}\, (\phi_2^*\, \phi_2)\,(\phi^*\,
\phi)^\frac{d_u}{2}\, ,
 \label{potential}
\end{eqnarray}
where $\mu$ is the parameter inserted in order to make the
couplings $\lambda_2$ and $\lambda^\prime_2$ dimensionless. Here,
our aim is to find the minimum of the potential V along the ray
$\Phi_{i}=\rho\, N_i$ with $\Phi_{i}=(\Phi_{1}, \Phi_{2}, \phi)$
(see \cite{Weinberg}) and in unitary gauge we have
\begin{eqnarray}
\Phi_{1}=\frac{\rho}{\sqrt{2}}\left(\begin{array}{c c}
0\\N_0\end{array}\right) \,\, ;
\phi_{2}=\frac{\rho}{\sqrt{2}}\,N^\prime_0 \,\,;
\phi=\frac{\rho}{\sqrt{2}}\,N_1 . \label{Phi12phi}
\end{eqnarray}
Since $\vec{N}$ is taken as the unit vector in the field space we
have
\begin{eqnarray}
N_0^2+N_0^{\prime 2}+N_1^2=1\, .\label{normalization}
\end{eqnarray}
Using the eqs. (\ref{potential}) and (\ref{Phi12phi}) we get
\begin{eqnarray}
V(\rho,N_i)&=& \frac{\rho^4}{4}\,\Bigg( \lambda_0\, N_0^4
+\lambda^\prime_0\, N^{\prime\, 4}_0 + 2\,\Big(
\frac{\hat{\rho}^2}{2}
\Big)^{-\epsilon}\,(\lambda_2\,N_0^2+\lambda^\prime_2\,N^{\prime\,
2}_0)\,N_1^{d_u}+\lambda_1\,N_1^4 \Bigg)\,. \label{potential2}
\end{eqnarray}
The stationary condition for the potential V, $\frac{\partial V}{
\partial N_i}|_{\vec{n}}$ along a special $\vec{n}$ direction reads
\begin{eqnarray}
\Big(\frac{\hat{\rho}^2}{2}\Big)^{-\epsilon}\,\lambda_2\,n_1^{d_u}&=&-\lambda_0\,
n_0^2 \,, \nonumber \\
\Big(\frac{\hat{\rho}^2}{2}\Big)^{-\epsilon}\,\lambda^\prime_2\,n_1^{d_u}&=&
-\lambda^\prime_0\, n^{\prime\, ^2}_0 \,, \nonumber \\
2\,\lambda_1\,n_1^4&=&d_u\,(\lambda_0\,n_0^4+\lambda^\prime_0\,
n_0^{\prime\,4}) \,, \label{n01d}
\end{eqnarray}
where $\epsilon=\frac{2-d_u}{2}$ and
$\hat{\rho}=\frac{\rho}{\mu}$. This condition and
eq.(\ref{normalization}) results in
\begin{eqnarray}
n^2_0&=&\frac{\chi}{1+\chi+\kappa} \,, \nonumber \\
n^{\prime\, ^2}_0&=&\frac{1}{1+\chi+\kappa} \,, \nonumber \\
n^2_1&=&\frac{\kappa}{1+\chi+\kappa} \, , \label{n01}
\end{eqnarray}
where
\begin{eqnarray}
\chi=\frac{\lambda^\prime_0\,\lambda_2}{\lambda_0\,\lambda^\prime_2}\, ,
 \nonumber \\
\kappa=\sqrt{\frac{d_u\,\chi\,\lambda_0\,(\lambda_2\,\chi+\lambda^\prime_2)}
{2\,\lambda_1\,\lambda_2}}  \, . \label{chikappa}
\end{eqnarray}
By using eq.(\ref{n01}) the nontrivial minimum value of the
potential is obtained as
\begin{eqnarray}
V(\rho, n_i)&=& -\frac{\rho^4}{4}\,\Bigg( \lambda_0\, n_0^4
+\lambda^\prime_0\, n^{\prime\, 4}_0  \Bigg)\,\epsilon\,.
\label{minpotential2}
\end{eqnarray}
Notice that, for $d_u=2$, the minimum of the potential is trivial,
namely $V(\rho, n_i)=0$. However, the nontrivial minimum can be
obtained at tree level for $1< d_u <2$ without the need for CW
mechanism (see \cite{Weinberg} for details of CW mechanism). The
stationary condition fixes the parameter $\rho$ as,
\begin{eqnarray}
\rho=\rho_0=\Bigg
(\frac{-2^\epsilon\,\lambda_2\,n_1^d}{\lambda_0\,n_0^2}
 \Bigg)^\frac{1}{2\,\epsilon}\,\mu
\, ,\label{rho1}
\end{eqnarray}
or
\begin{eqnarray}
\rho=\rho_0=\Bigg
(\frac{-2^\epsilon\,\lambda^\prime_2\,n_1^d}{\lambda^\prime_0\,n^{\prime\,2}_0}
 \Bigg)^\frac{1}{2\,\epsilon}\,\mu
\, ,\label{rho1}
\end{eqnarray}
and this parameter should be stabilized as $d_u\rightarrow 2$,
since it is responsible for the mass scales of the theory. When
one chooses $d_u=2$ in the stationary conditions (see eq.
(\ref{n01d})), the restriction which connects the couplings
$\lambda_0$, $\lambda^\prime_0$, $\lambda_2$ and
$\lambda^\prime_2$ can be reached naturally:
\begin{eqnarray}
\sqrt{\lambda_0\,\lambda^\prime_0\,\lambda_1}=-\sqrt{\lambda^\prime_0\,
\lambda_2^2+ \lambda_0\,\lambda^{\prime\,2}_2}\, ,\label{restr}
\end{eqnarray}
and this is the  sufficient condition in order to stabilize the
parameter $\rho_0$. If we consider this restriction we get
\begin{eqnarray}
\hat{\rho_0}^2=2\,\Big(\frac{d}{2}\Big)^{\frac{d}{4-2\,d}}\,\Big(\frac{-\lambda^\prime_2}
{\lambda^\prime_0}\Big)\,\Big(1-\sqrt{\frac{d}{2}}\,\frac{\lambda^\prime_0}
{\lambda^\prime_2}+\frac{\lambda^\prime_0\,\lambda_2}
{\lambda^\prime_2\,\lambda_0}\Big) \, ,\label{rhorestr}
\end{eqnarray}
and, for $d_u\rightarrow 2$, it converges to
\begin{eqnarray}
\hat{\rho}^2_0\rightarrow
2\,\frac{\lambda^\prime_0\,(-\lambda_2)+(-\lambda^\prime_2)\,\lambda_0+
\lambda_0\,\lambda^\prime_0}
{\sqrt{e}\,\lambda_0\,\lambda^\prime_0}
 \, .\label{rhorestrdu2}
\end{eqnarray}
At this stage we consider that the couplings $\lambda_2$,
$\lambda^\prime_2$ which drive the strengths of the neutral Higgs
boson-the massless scalar field vertices have the same strength.
With this choice the constraint eq. (\ref{restr}) becomes
\begin{eqnarray}
\lambda_2=-\sqrt{\frac{\lambda_0\,\lambda^\prime_0\,\lambda_1}
{\lambda_0+\lambda^\prime_0 }} \, ,\label{restr2}
\end{eqnarray}
and parameters $\chi$ and $\kappa$ read
\begin{eqnarray}
\chi=\frac{\lambda^\prime_0}{\lambda_0}\, ,
\nonumber \\
\kappa=\sqrt{\frac{d_u}{2}}\,\sqrt{\frac{\lambda^\prime_0\,
(\lambda_0+\lambda^\prime_0)} {\lambda_0\,\lambda_1}} \, .
\label{chikappa2}
\end{eqnarray}

Now we are ready to study the mixing matrix of the scalars under
consideration. If one expands the fields $\Phi_{1}$, $\phi_{2}$
and $\phi$ around the vacuum as
\begin{eqnarray}
\Phi_{1}=\frac{1}{\sqrt{2}}\left(\begin{array}{c c}
0\\\rho_0\,n_0+h\end{array}\right) \,\, ;
\phi_{2}=\frac{1}{\sqrt{2}}\,(\rho_0\,n^\prime_0+h^\prime) \,\,;
\phi=\frac{1}{\sqrt{2}}\,(\rho_0\,n_1+s) , \label{Phi12phi2}
\end{eqnarray}
the potential (eq. (\ref{potential})) reads
\begin{eqnarray}
V(h, h^\prime, s)\!\!\!\!\!&=&\!\!\!\!\! \frac{\lambda_0}{4}\,
(\rho_0\,n_0+h)^4+ \frac{\lambda^\prime_0}{4}\,
(\rho_0\,n^\prime_0+h^\prime)^4+\frac{\lambda_1}{4}\,
(\rho_0\,n_1+s)^4 + 2^{-\frac{d}{2}}\,
\lambda_2\,\mu^{2\,\epsilon}\,
(\rho_0\,n_0+h)^2\,(\rho_0\,n_1+s)^{d_u}\nonumber \\
&+&2^{-\frac{d}{2}}\, \lambda^\prime_2\,\mu^{2\,\epsilon}\,
(\rho_0\,n^\prime_0+h^\prime)^2\,(\rho_0\,n_1+s)^{d_u}\, .
\label{potential3}
\end{eqnarray}
Using this potential we get the mass matrix
$(M^2)_{ij}=\frac{\partial^2\,V}{\partial\,\phi_i\,\partial\,\phi_j}|_{\phi_i=0}$
with $\phi_i=(h,h^\prime,s)$ as
\\
\begin{eqnarray}
(M^2)_{ij}=2\,\rho_0^2\,n_0^2\, \left(%
\begin{array}{ccc}
\lambda_0 & 0 & -\Big(\frac{d_u\,\lambda_0}{2} \Big)^\frac{3}{4}\,
\Big(\frac{\lambda^\prime_0\,\lambda_1}
{\lambda_0+\lambda^\prime_0} \Big)^\frac{1}{4} \\ \\0 & \lambda_0
& -\Big(\frac{d_u\,\lambda_0}{2} \Big)^\frac{3}{4}\,
\Big(\frac{\lambda_0^2\,\lambda_1}
{\lambda^\prime_0\,(\lambda_0+\lambda^\prime_0)} \Big)^\frac{1}{4}
\\ \\
 -\Big(\frac{d_u\,\lambda_0}{2} \Big)^\frac{3}{4}\,
\Big(\frac{\lambda^\prime_0\,\lambda_1}
{\lambda_0+\lambda^\prime_0} \Big)^\frac{1}{4} &
-\Big(\frac{d_u\,\lambda_0}{2}\Big)^\frac{3}{4}\,
\Big(\frac{\lambda_0^2\,\lambda_1}
{\lambda^\prime_0\,(\lambda_0+\lambda^\prime_0)} \Big)^\frac{1}{4}
& (2-\frac{d_u}{2})\,
\sqrt{\frac{d_u}{2}}\,\sqrt{\frac{\lambda_0\,\lambda_1\,
(\lambda_0+\lambda^\prime_0)}{\lambda^\prime_0}}\,, \\
\end{array}%
\right). \label{restr2}
\end{eqnarray}
The eigenvalues of the matrix are
\begin{eqnarray}
m_I^2&=& 2\,\lambda_0\,n_0^2\,\rho_0^2\, , \nonumber \\
m_{II}^2&=& \lambda_0\,n_0^2\,\rho_0^2\, \Bigg(
1+(2-\frac{d_u}{2})\,\sqrt{\frac{d_u\,s_{10}\,(1+s_0)}{2\,s_0}}-
\sqrt{\Delta}
\Bigg)\, , \nonumber \\
m_{III}^2&=& \lambda_0\,n_0^2\,\rho_0^2\, \Bigg(
1+(2-\frac{d_u}{2})\,\sqrt{\frac{d_u\,s_{10}\,(1+s_0)}{2\,s_0}}+
\sqrt{\Delta} \Bigg)\, ,\label{mass2}
\end{eqnarray}
where
\begin{eqnarray}
\Delta=d_u\,\sqrt{\frac{2 d_u\,s_{10}\,(1+s_0)}{s_0}}+\Bigg(
 1+(\frac{d_u}{2}-2)\,\sqrt{\frac{d_u\,s_{10}\,(1+s_0)}{2\,s_0}}\Bigg)^2
 \,.
 \label{delta}
\end{eqnarray}
Here we used the parametrization
\begin{eqnarray}
\lambda^\prime_0&=&s_0\,\lambda_0\, ,
\nonumber \\
\lambda_1&=&s_{10}\,\lambda_0\  \, . \label{lambda01}
\end{eqnarray}
The physical states $h_I,\,h_{II},\,h_{III}$ are connected to the
original states $h,\,h^\prime,\,s$ as
\begin{eqnarray}
\left(%
\begin{array}{c}
  h \\
  h^\prime \\
  s \\
\end{array}%
\right)=\left(%
\begin{array}{ccc}
  c_\alpha & -c_\eta\,s_\alpha & s_\eta\,s_\alpha \\
  s_\alpha & c_\eta\,c_\alpha & -s_\eta\,c_\alpha \\
  0 & s_\eta & c_\eta \\
\end{array}%
\right)
\left(%
\begin{array}{c}
  h_I \\
  h_{II} \\
  h_{III} \\
\end{array}%
\right)\,, \label{diagMatrix}
\end{eqnarray}
where $c_{\alpha\,(\eta)}=cos\,\alpha\,(\eta)$,
$s_{\alpha\,(\eta)}=sin\,\alpha\,(\eta)$ and
\begin{eqnarray}
tan\,2\,\alpha&=& \frac{2\,\sqrt{s_0}}{s_0-1}\,, \nonumber
\\
tan\,2\,\eta&=& \Big(\frac{d_u}{2}\Big)^\frac{3}{4}\,\frac{2
\Big(s_0\,s_{10}\,(1+s_0)\,\Big)^\frac{1}{4}}{(1-\frac{d_u}{4})\,
\sqrt{2\,d_u\,s_{10}\,(1+s_0)}-\sqrt{s_0}}\, .
  \,\label{tan2alfeta}
\end{eqnarray}
When $d_u\rightarrow 2$, the state $h_{II}$ is massless in the
tree level and it has the lightest mass for $1< d_u< 2$. $h_{I}$
and $h_{III}$ can be identified as the SM Higgs boson and heavy
scalar coming from the shadow sector, respectively.

Finally, we construct the last restriction by fixing the vacuum
expectation value $v_0=n_0\,\rho_0$, by the gauge boson mass $m_W$
as
\begin{eqnarray}
v_0^2=\frac{4\,m_W^2}{g_W^2}=\frac{1}{\sqrt{2}\,G_F}\, ,
\label{v02}
\end{eqnarray}
where $G_F$ is the Fermi constant. By using eqs. (\ref{n01}) and
(\ref{rhorestr}) we get
\begin{eqnarray}
\hat{v}_0^2=c_0\, \frac{s_{10}\,\sqrt{2\, s_0
\,(1+s_0)}+s_{0}\,\sqrt{d_u\,s_{10}}} {\sqrt{d\, s_0
\,(1+s_0)}+(1+s_0)\,\sqrt{2\,s_{10}}} \, , \label{hatv02}
\end{eqnarray}
with $c_0=2\,\Big(\frac{d_u}{2}\Big)^\frac{d_u}{2\,(2-\,d_u)}$.
The choice of the parameter $\mu$ around weak scale as $\mu=v_0$
results in additional restriction which connects parameters $s_0$
and $s_{10}$ (see eq. (\ref{hatv02}) by considering
$\hat{v}_0^2=1$) as
\begin{eqnarray}
s_{10}=\frac{1+s_0}{c_0^2\,s_0}  \, . \label{s10}
\end{eqnarray}
When $d_u\rightarrow 2$, $s_{10}\rightarrow
\frac{e}{4}\,\frac{1+s_0}{s_0}$ and when $d_u\rightarrow 1$,
$s_{10}\rightarrow \frac{1+s_0}{2\,s_0}$. It is shown that the
ratios are of the order of one and the choice $\mu=v_0$ is
reasonable (see \cite{JPLee} for the similar discussion.)
\\ \\
{\Large \textbf{The effect of the unparticle-neutral scalars
mixing on the muon anomalous magnetic moment}}
\\ \\
%\section{The effect of the unparticle-neutral scalars mixing on the anomalous
%magnetic moment of the muon in this toy  model.}
%
The current experimental world average of the muon AMM by the
latest BNL experiment \cite{BNL} has been announced as
\begin{eqnarray}
a_{\mu}=116\, 592\, 080\, (63)\times 10^{-11}\,\, .
\end{eqnarray}
From the theoretical point of view,  the muon AMM is written in
terms of different contributions in the framework of the SM as
\begin{eqnarray}
a_{\mu}(SM)=a_{\mu}(QED)+a_{\mu}(weak)+a_{\mu}(hadronic) \,\, ,
\end{eqnarray}
where $a_{\mu}(QED)=116\,\, 584\,\,718.09\,(0.14)\,(0.04)\times
10^{-11}$ and $a_{\mu}(weak)=152\,\,(2)\,(1)\times 10^{-11}$. The
hadronic contributions are under theoretical investigation and
need the forthcoming results from the high precision measurements.
With the new data from Novosibirsk, some exclusive channels from
BaBar and the compilation of  the $e^+ e^- $ data by Michel Davier
and collaborators the numerical value
$6908.7\,(39)\,(19)\,(7)\times 10^{-11}$ (see \cite{Rafael},
\cite{Zhang} and references therein) is obtained for the leading
order hadronic vacuum polarization to $a_{\mu}(SM)$. The higher
order contribution (next to leading contributions) is estimated as
$-97.9\,(0.9)\,(0.3)\times 10^{-11}$ (see \cite{Calmet,Brodsky,
Lautrup, Barbieri, TTeubner}) and light by light scattering is
calculated as $105\,(26)\times 10^{-11}$ (see \cite{Knecht1,
Knecht2, Hayakawa, Prades, Melnikov, Rafael2}). Finally, one gets
the SM result
\begin{eqnarray}
a_{\mu}(SM)=116\, 591\, 785\, (51)\times 10^{-11}\, ,
\end{eqnarray}
which shows that there is still $3.6\,\sigma$ discrepancy between
the experimental result and the SM one. Now, we will study the
additional effect on the muon AMM due to the unparticle-neutral
scalars mixing system of our toy model and check whether it is
possible to explain the present discrepancy ($\delta$) by playing
with the free parameters of the model.  By considering the highest
and the lowest experimental and SM values of the muon AMM,
$\delta$ is estimated in the range
\begin{eqnarray}
1.810\times 10^{-9} < \delta < 4.089\times 10^{-9} \, .
\end{eqnarray}
Here, we will check if the new contribution coming from the mixing
spectrum\footnote{In the calculation of upper and lower limits of
$\delta$ we take the SM electroweak contribution without the SM
Higgs one, since we insert our mixing spectrum effect instead. We
see that the SM Higgs  effect is of the order of $10^{-12}$.}
reaches to the values of the order of $10^{-9}$.

The starting point is the SM Higgs-quark (lepton) interaction in
the SM
\begin{eqnarray}
{\cal{L}}_{Y}&=&\eta^{U}_{ij}\, \bar{Q}_{i L}\, \tilde{\Phi}_{1}\,
U_{j R}+ \eta^{D(E)}_{ij}\,  \bar{Q}_{i L}\, (\bar{l}_{i L})\,
\phi_{1}\,  D_{j R} \,(E_{j R})+ h.c. \,\,\, \label{lagrangian}
\end{eqnarray}
where $L=\frac{1}{2}(1-\gamma_5)$ and $R=\frac{1}{2}(1+\gamma_5)$
denote chiral projections, $\Phi_{1}$ is the scalar doublet, $Q_{i
L}\,(l_{i L})$ for $i=1,2$, are quark (lepton) doublets, and $U_{j
R}$, $D_{j R}$ ($E_{i R}$) are right, handed up type, down type
quark singlets (right handed lepton singlets), $\eta^{U,D,
E}_{ij}$ is the matrices of the Yukawa couplings.

\begin{figure}[htb]
\vskip 7.5truein \centering \epsfxsize=5.0in
\leavevmode\epsffile{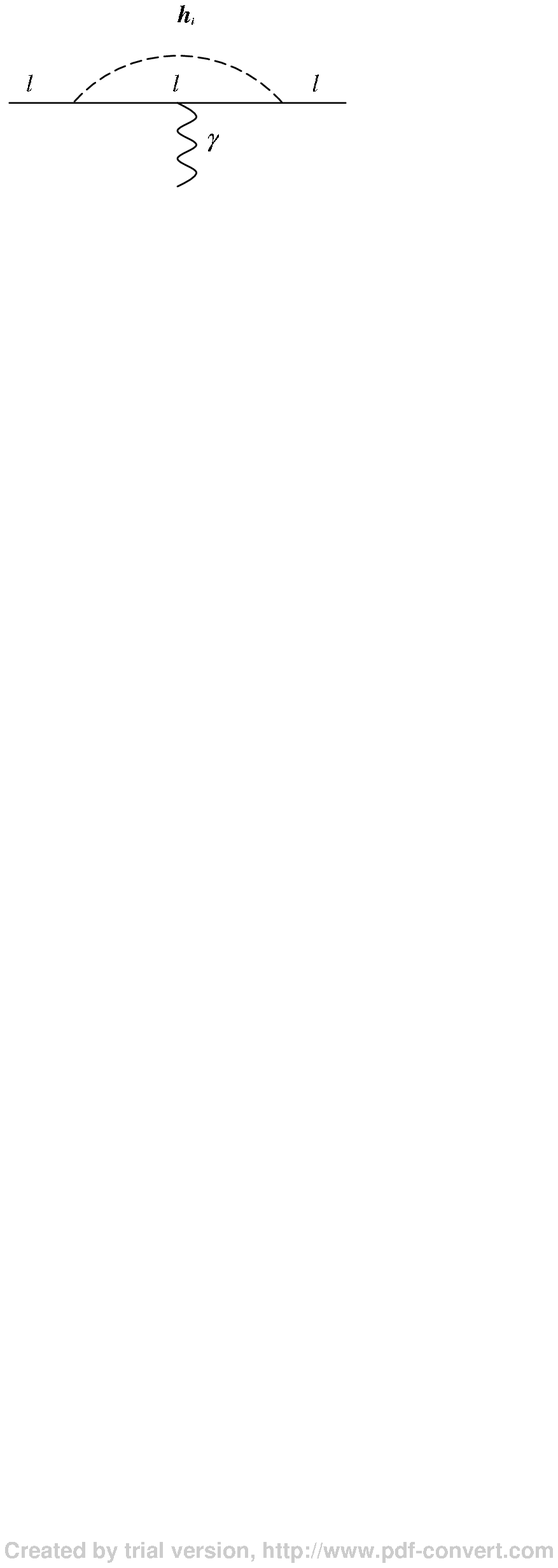} \vskip -13.5truein \caption[]{One
loop diagrams contributing to AMM of $l$-lepton  due to the
neutral scalars $h_i$ where $i=I,II,III$. Wavy (dashed) line
represents the electromagnetic field ($h_i$ fields).} \label{fig1}
\end{figure}

Our aim is to estimate the effect of the unparticle-neutral
scalars mixing on the muon AMM and, therefore, we consider the
charged leptons in our calculations. The charged lepton masses are
proportional to the vacuum expectation value $v_0=n_0\,\rho_0$ and
after the spontaneous breakdown of the electroweak symmetry the
$i^{th}$ charged lepton mass is obtained as $m_i=v_0\, \eta_{ii}$,
with $v_0=\frac{2\,m_W}{g_W}$ and, therefore, the coupling
$\eta_{ii}$ is $\eta_{ii}=\frac{m_i}{v_0}$.
\newpage
The effective interaction\footnote{The most general
Lorentz-invariant form of the coupling of a charged lepton to a
photon of four-momentum $q_{\nu}$ can be written as
$ \Gamma_{\mu}= G_1 (q^2)\, \gamma_{\mu} + G_2 (q^2)\,
\sigma_{\mu\nu} \,q^{\nu} + G_3 (q^2)\, \sigma_{\mu\nu}\gamma_5\,
q^{\nu} $
where $q_{\nu}$ is the photon 4-vector and the $q^2$ dependent
form factors $G_{1}(q^2)$, $G_{2}(q^2)$ and $G_{3}(q^2)$ are
proportional to the charge, AMM and EDM of the $l$-lepton
respectively.} for the anomalous magnetic moment of the lepton is
defined as
\begin{eqnarray}
{\cal L}_{AMM}=a_l \frac{e}{4\,m_l}\,\bar{l}
\,\sigma_{\mu\nu}\,l\,
 F^{\mu\nu} \,\, , \label{AMM1}
\end{eqnarray}
where $F_{\mu\nu}$ is the electromagnetic field tensor and "$a_l$"
is the AMM of the lepton "$l$", $(l=e,\,\mu,\,\tau)$. This
interaction can be induced by the unparticle-neutral scalars
mixing system  at loop level in the toy model under
consideration\footnote{Here we do not take charged FC interaction
in the leptonic sector due to the small couplings for $\mu-\nu_l$
transitions.} (see Fig. \ref{fig1}). Since charged leptons couple
to the SM Higgs doublet, in the internal line, only the neutral
Higgs $h$ and, therefore, the physical states
$h_I,\,h_{II},\,h_{III}$ appear since
\begin{eqnarray}
h=c_\alpha\,h_I-c_\eta\,s_\alpha\,h_{II}+s_\eta\,s_\alpha\,h_{III}
\,\, , \label{h1Exp}
\end{eqnarray}
where $h_I,\,h_{II}$ and $h_{III}$ are identified as new neutral
scalar boson, scalon and the SM Higgs boson (see eqs.
(\ref{diagMatrix}) and (\ref{tan2alfeta}) for the physical states
and the parameters $c_\alpha$, $c_\eta$, $s_\alpha$ and $s_\eta$).
Using the definition of AMM of the lepton $l$ (eq. (\ref{AMM1})),
we get
\begin{eqnarray}
a^{New}_{\mu}=a^{h_I}_{\mu}+a^{h_{II}}_{\mu}+a^{h_{III}}_{\mu}
\,\, , \label{muNOM}
\end{eqnarray}
where
\begin{eqnarray}
a^{h_I}_{\mu}&=&\frac{G_F\,m_\mu^4\,c^2_\alpha}{4\,\sqrt{2}\,\pi^2\,
m^2_{h_I}}\,\int_0^1\,dx\,\frac{(2-x)\,x^2}
{1+x\,(x\,\frac{m^2_\mu}{m^2_{h_I}}-1)} \nonumber\, , \\
a^{h_{II}}_{\mu}&=&\frac{G_F\,m_\mu^4\,s^2_\alpha\,c^2_\eta}{4\,\sqrt{2}
\,\pi^2\,m^2_{h_{II}}}\,\int_0^1\,dx\,\frac{(2-x)\,x^2}
{1+x\,(x\,\frac{m^2_\mu}{m^2_{h_{II}}}-1)} \nonumber\, , \\
a^{h_{III}}_{\mu}&=&\frac{G_F\,m_\mu^4\,s^2_\alpha\,s^2_\eta}{4\,\sqrt{2}\,
\pi^2\,m^2_{h_{III}}}\,\int_0^1\,dx\,\frac{(2-x)\,x^2}
{1+x\,(x\,\frac{m^2_\mu}{m^2_{h_{III}}}-1)} \, . \label{tauANOM}
\end{eqnarray}
%
%%%
%%%
{\Large \textbf{Discussion}}
\\
%\section{Discussion}
%

In the present work  we extend the scalar sector by considering a
shadow Higgs one with complex scalar and choose a scalar potential
responsible for the mixing of neutral scalars. Here, the
motivation is to drive the electroweak symmetry breaking at tree
level and, for this, the hidden sector is chosen as the unparticle
sector propsed by Georgi \cite{Georgi1} recently. The unparticle
sector causes the electroweak symmetry breaking without need for
CW mechanism (see \cite{JPLee}) since this sector is scale
invariant and the unparticle operator has non-integral scaling
dimension. When the scaling dimension reaches to $d_u=2$ one gets
the trivial minimum and the CW mechanism needs for the electroweak
symmetry breaking. On the other hand, when the electroweak
symmetry is broken, the scalar fields are mixed and three massive
states appear. Therefore, the hidden sector scale invariance is
also broken.

There are number of parameters  existing in our toy model and they
should be restricted by conditions based on mathematical and
physical backgrounds. At first, we choose that the couplings
$\lambda_2$ and $\lambda^\prime_2$ which drive the strengths of
the neutral Higgs boson-the massless scalar field vertices have
the same strength. Second, we construct the
restriction\footnote{This restriction is sufficient to stabilize
the parameter $\rho_0$ which plays the role of mass scale of the
theory.} which connects the couplings $\lambda_0$,
$\lambda^\prime_0$, $\lambda_2$ naturally (see
eq.({\ref{restr}})). The choice of $\mu$ around weak scale, namely
$\mu=v_0$, results in a new restriction and the couplings
$\lambda_1$ and $\lambda_0$ are also connected (see
eq.(\ref{s10})). Finally, we have the couplings $\lambda_0$,
$\lambda^\prime_0$, and the scaling dimension $d_u$ as free
parameters. Notice that we take the state $h_{I}$ as the SM Higgs
boson and choose its mass $m_{I}$ as a fixed value, which will be
hopefully measured as a SM Higgs mass in the forthcoming
experiments.

\begin{figure}[htb]
\vskip -3.0truein \centering \epsfxsize=6.8in
\leavevmode\epsffile{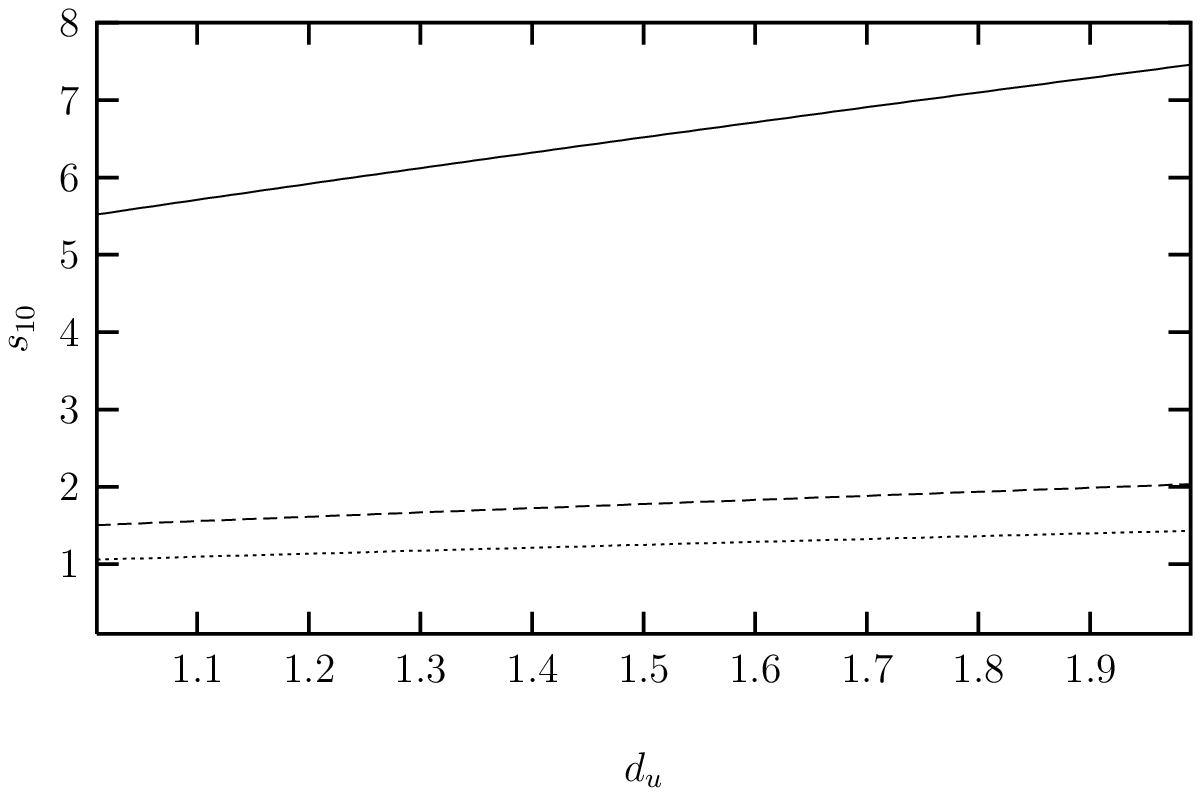} \vskip -3.0truein
\caption[]{$s_{10}$ as a function of $d_u$. Here the solid
(dashed, dotted) line represents $s_{10}$ for $s_0=0.1$
($s_0=0.5$, $s_0=0.9$).} \label{s10du}
\end{figure}

In Fig.\ref{s10du} we plot $d_u$ dependence of $s_{10}$ for
different values of the parameter
$s_0=\frac{\lambda^\prime_0}{\lambda_0}$. Here the solid (dashed,
dotted) line represents $s_{10}$ for $s_0=0.1$ ($s_0=0.5$,
$s_0=0.9$). $s_{10}$ is  sensitive to $d_u$, especially for small
$s_0$, and increases with the decreasing values of $s_0$ and the
increasing values of $d_u$.

\begin{figure}[htb]
\vskip -3.0truein \centering \epsfxsize=6.8in
\leavevmode\epsffile{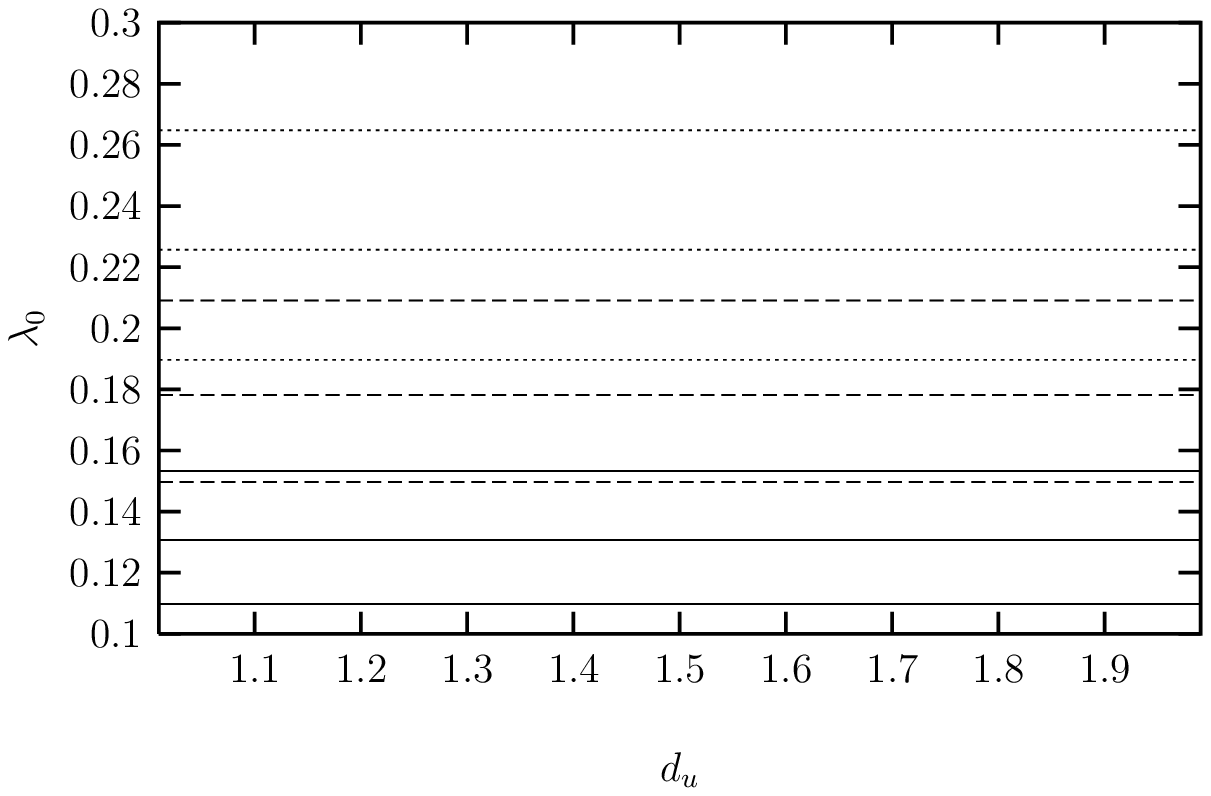} \vskip -3.0truein
\caption[]{$\lambda_0$ as a function of $d_u$. Here the
lower-intermediate-upper solid (dashed, dotted) line represents
$\lambda_0$ for $m_{I}=110-120-130$ (GeV), $s_0=0.1$ ($s_0=0.5$,
$s_0=0.9$).} \label{lam0duB}
\end{figure}

Fig.\ref{lam0duB} represents $d_u$ dependence of $\lambda_0$ for
different values of the parameter $s_0$ and the mass $m_{I}$. Here
the lower-intermediate-upper solid (dashed, dotted) line
represents $\lambda_0$ for $m_{I}=110-120-130$ (GeV), $s_0=0.1$
($s_0=0.5$, $s_0=0.9$). We observe that $\lambda_0$ does not
depend on  the scaling dimension $d_u$ and the increasing values
of the mass $m_{I}$ result in the enhancement of the coupling
$\lambda_0$. On the other hand $\lambda_0$ increases with the
increasing values of $s_0$.

\begin{figure}[htb]
\vskip -3.0truein \centering \epsfxsize=6.8in
\leavevmode\epsffile{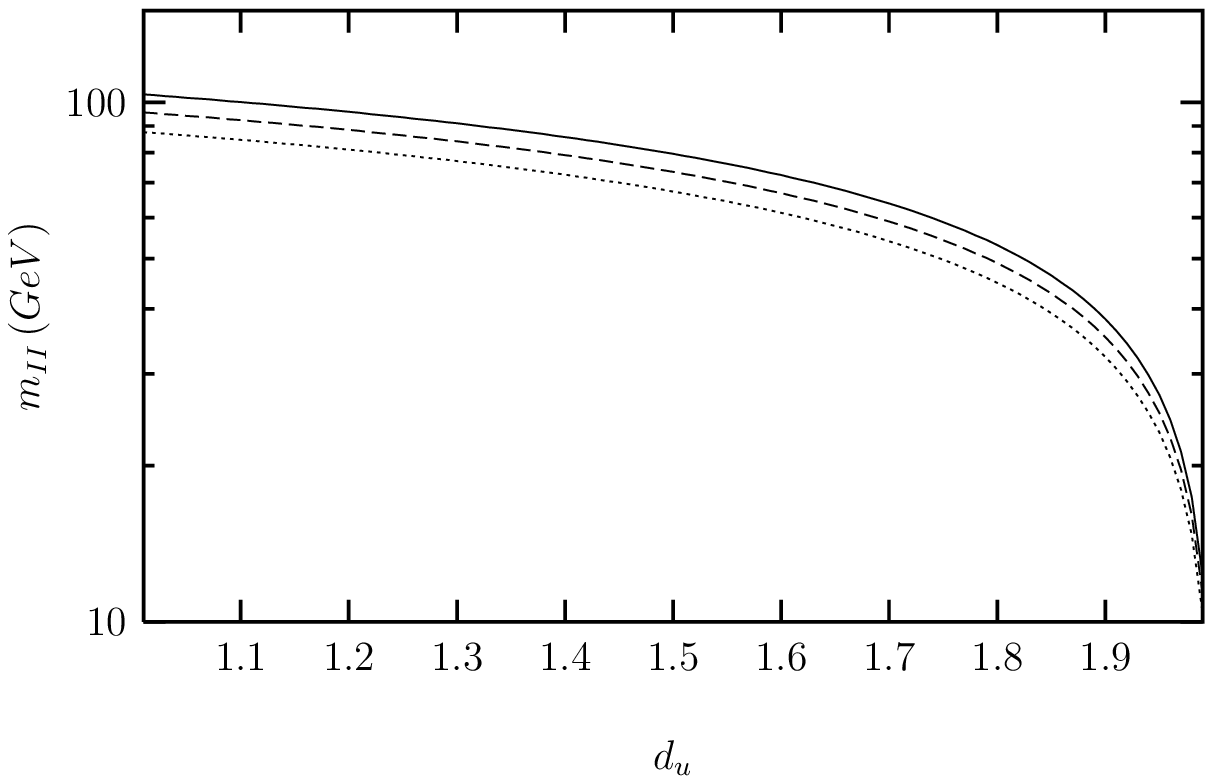} \vskip -3.0truein
\caption[]{$m_{II}$ as a function of $d_u$ for $s_0=0.1$. Here the
solid (dashed, dotted) line represents $m_{II}$ for
$m_{I}=130-120-110$ (GeV).} \label{m2du}
\end{figure}

\begin{figure}[htb]
\vskip -3.0truein \centering \epsfxsize=6.8in
\leavevmode\epsffile{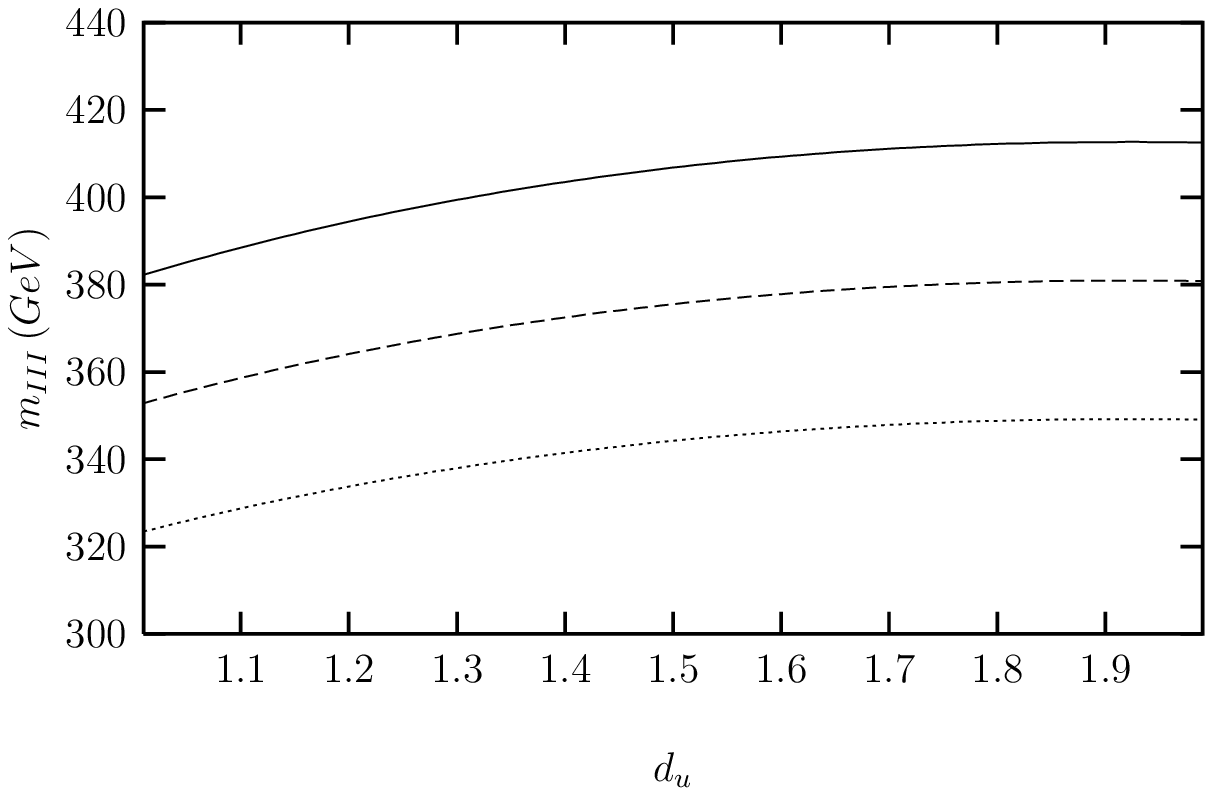} \vskip -3.0truein
\caption[]{$m_{III}$ as a function of $d_u$ for $s_0=0.1$. Here
the solid (dashed, dotted) line represents $m_{III}$ for
$m_{I}=130-120-110$ (GeV).} \label{m3du}
\end{figure}

In Figs. \ref{m2du} and \ref{m3du} we show $d_u$ dependence of the
masses $m_{II}$ and $m_{III}$ for different values of the mass
$m_{III}$. Fig.\ref{m2du} is devoted to $d_u$ dependence of the
mass $m_{I}$ for the parameter $s_0=0.1$. Here the solid (dashed,
dotted) line represents $m_{II}$ for $m_{I}=130-120-110$ (GeV). We
see that $m_{II}$ reaches to zero when $d_u=2$ at tree level and
this is the case that $h_{II}$ is the pseudo Golstone boson due to
the spontaneous symmetry breaking of conformal symmetry. With the
increasing values of the mass $m_{I}$, $m_{II}$ also increases.
Fig.\ref{m3du} shows $d_u$ dependence of the mass $m_{III}$ for
the parameter $s_0=0.1$. Here the solid (dashed, dotted) line
represents $m_{III}$ for $m_{I}=130-120-110$ (GeV). The state
$h_{III}$ is the heaviest one  and it is appropriate to expect
that it is the new neutral shadow scalar.

\begin{figure}[htb]
\vskip -3.0truein \centering \epsfxsize=6.8in
\leavevmode\epsffile{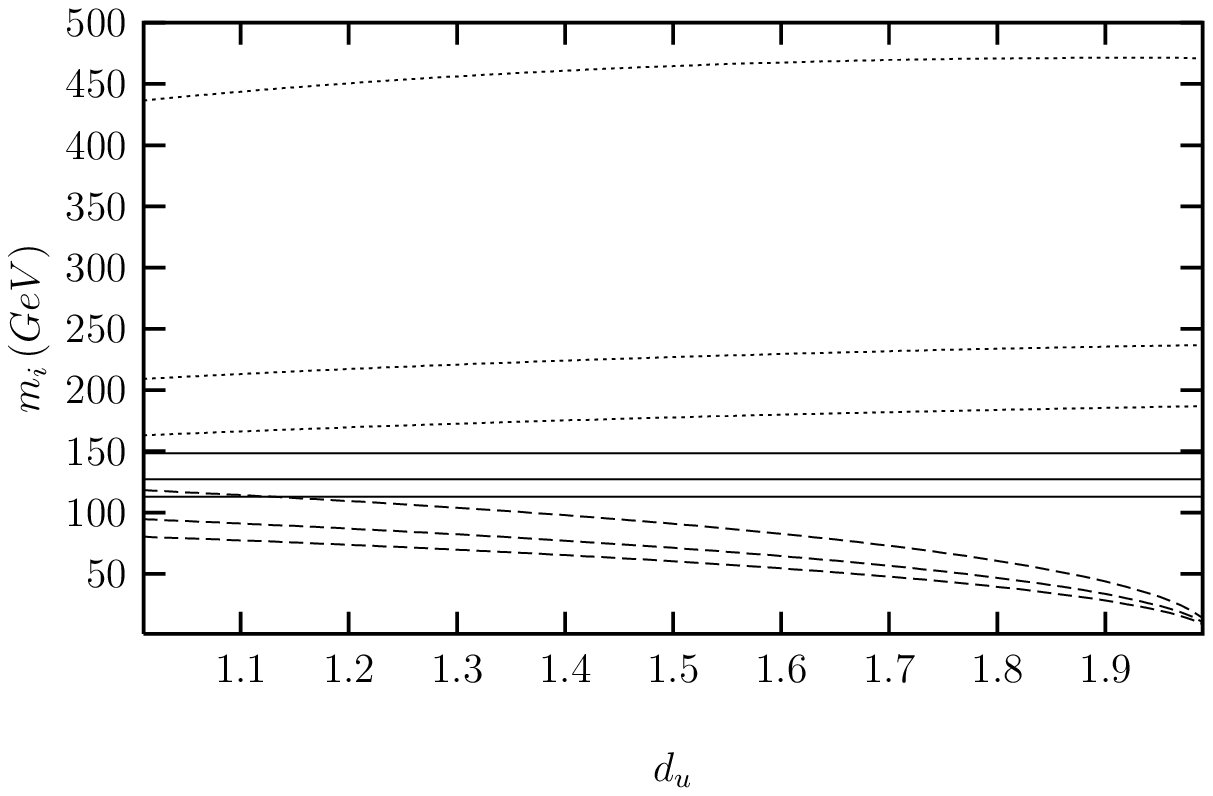} \vskip -3.0truein \caption[]{$m_i$
as a function of $d_u$ for $\lambda_0=0.2$. Here the
upper-intermediate-lower solid (dashed, dotted) line represents
$m_{I}$ ($m_{II}$, $m_{III}$) for $s_0=0.1-0.5-0.9$.}
\label{m123du}
\end{figure}

For completeness, in Fig  \ref{m123du}, we present $d_u$
dependence of the masses $m_{I}$, $m_{II}$ and $m_{III}$ without
restricting $m_{I}$. Here the upper-intermediate-lower solid
(dashed, dotted) line represents $m_{I}$ ($m_{II}$, $m_{III}$) for
$s_0=0.1-0.5-0.9$ and for $\lambda_0=0.2$. Here we observe that
$h_{III}$ is the heaviest state and $h_{II}$ becomes massless when
$d_u=2$ at tree level as expected. $m_{III}$ increases, $m_{II}$
is fixed and $m_{I}$ decreases with the increasing values of
$d_u$. Furthermore, the masses of the scalars increase with the
decreasing values of $s_0$.

Now, we start to analyze the effect of the unparticle-neutral
scalars mixing spectrum on the muon AMM and we study  $d_u$ and
$s_0$ dependence of the part of the muon AMM, $AMM_{scl}$, which
is driven by the mixed scalar spectrum.

\begin{figure}[htb]
\vskip -3.0truein \centering \epsfxsize=6.8in
\leavevmode\epsffile{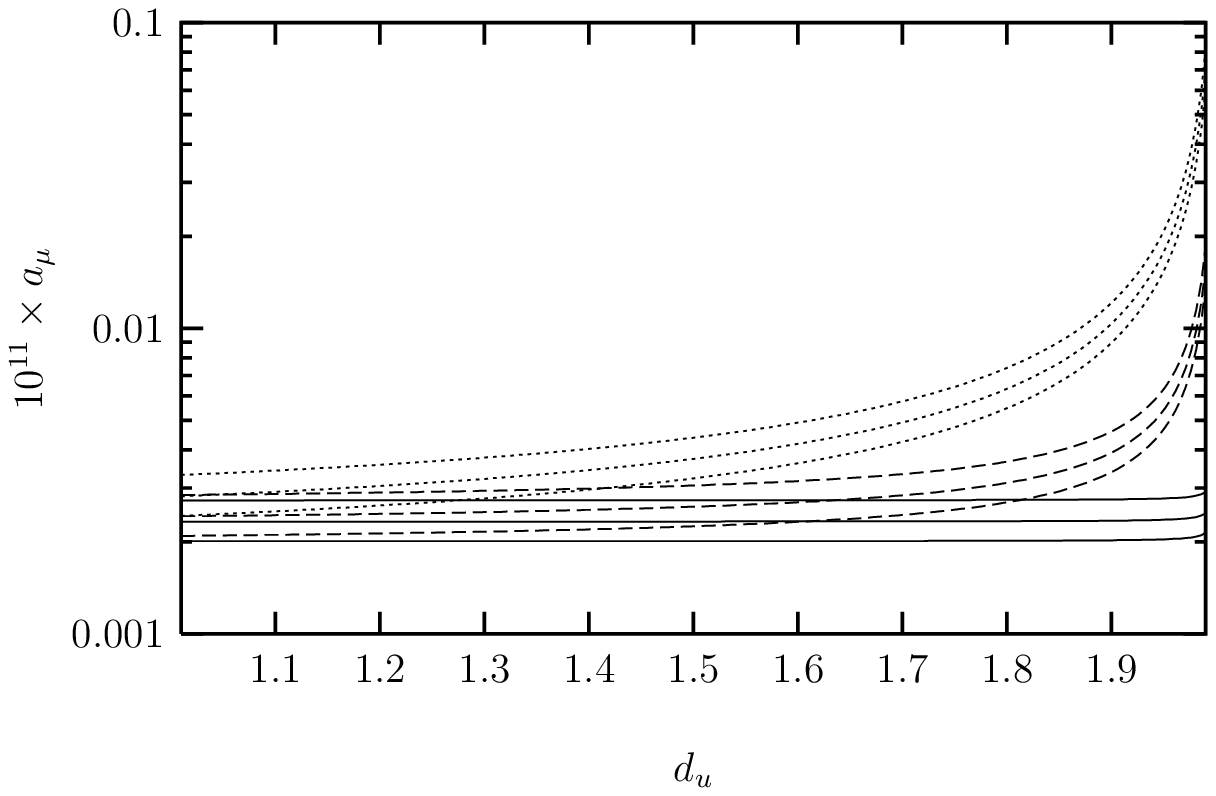} \vskip -3.0truein
\caption[]{$AMM_{scl}$ as a function of $d_u$. Here the
lower-intermediate-upper solid (dashed, dotted) line represents
$AMM_{scl}$ for $m_{I}=130-120-110$ (GeV), $s_0=0.001$ ($s_0=0.1$,
$s_0=0.9$). } \label{anomdu}
\end{figure}

Fig  \ref{anomdu} represents $d_u$ dependence of the $AMM_{scl}$
for different values of the parameter $s_0$ and the mass $m_{I}$.
Here the lower-intermediate-upper solid (dashed, dotted) line
represents $AMM_{scl}$ for $m_{I}=130-120-110$ (GeV), $s_0=0.001$
($s_0=0.1$, $s_0=0.9$). It is shown that $AMM_{scl}$ increases
with increasing values $s_0$ the scaling dimension $d_u$,
especially for large values of $s_0$. The $AMM_{scl}$ reaches to
the values of the order of $10^{-12}$ and it is almost three order
smaller than the discrepancy between the experimental and the SM
results of muon AMM.

\begin{figure}[htb]
\vskip -3.0truein \centering \epsfxsize=6.8in
\leavevmode\epsffile{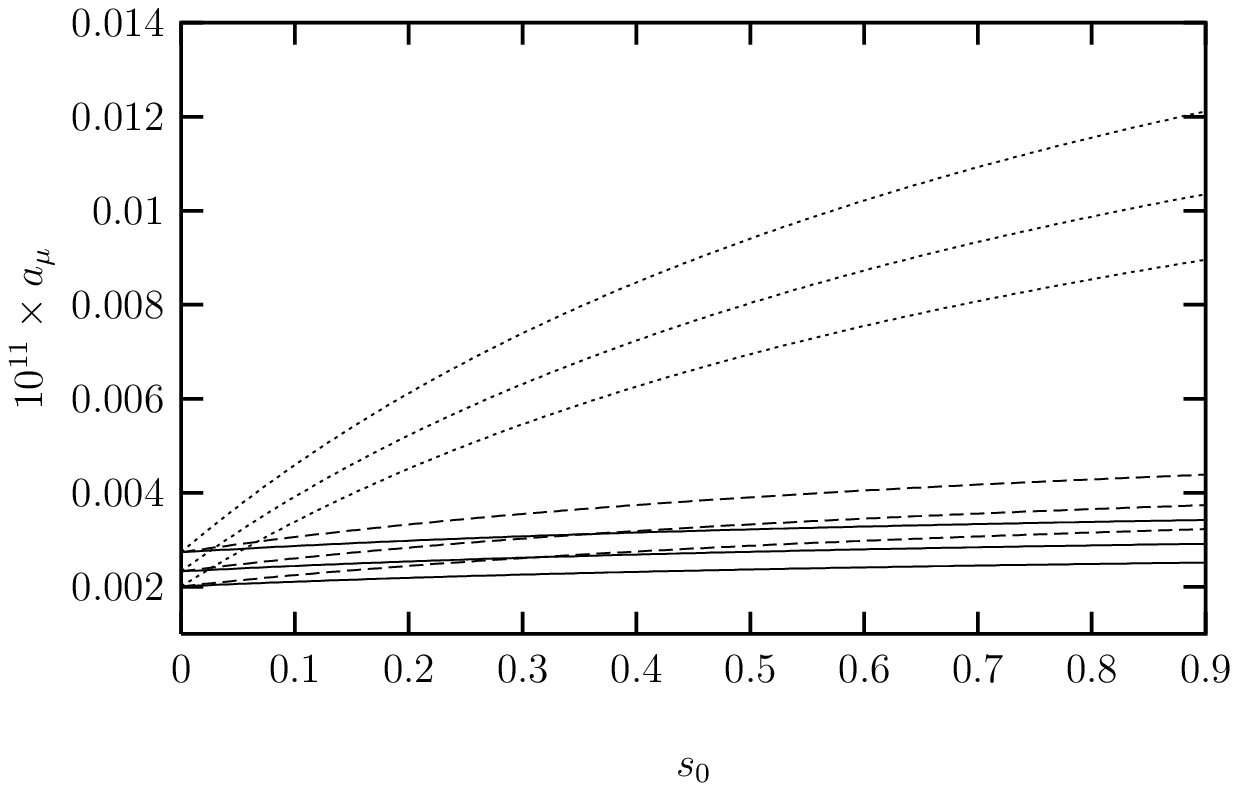} \vskip -3.0truein
\caption[]{$AMM_{scl}$ as a function of  $s_0$. Here the
lower-intermediate-upper solid (dashed, dotted) line represents
$AMM_{scl}$ for $m_{I}=130-120-110$ (GeV), $d_u=1.1$ ($d_u=1.5$,
$d_u=1.9$).} \label{anoms0}
\end{figure}

In Fig  \ref{anoms0} we show the $s_0$ dependence of the
$AMM_{scl}$ for different values of the parameter $d_u$ and the
mass $m_{I}$ Here the lower-intermediate-upper solid (dashed,
dotted) line represents $AMM_{scl}$ for $m_{I}=130-120-110$ (GeV),
$d_u=1.1$ ($d_u=1.5$, $d_u=1.9$). We observe that  the $AMM_{scl}$
is relatively sensitive to $s_0$ for its large values.

In summary, the interaction of the SM Higgs doublet with the
hidden unparticle sector is a possible candidate to drive the
electroweak symmetry breaking at tree level. We study this
breaking by considering the SM Higgs doublet and an additional
scalar and estimate the contribution of new scalar spectrum on the
muon AMM. We observe that this contribution is almost three order
less than the discrepancy between the experimental and the SM
results of muon AMM.
%%%
%%%

\newpage
\newpage
\end{document}